\documentclass[12pt]{article}

\usepackage{amsmath}
\usepackage{amssymb}
\usepackage{amsfonts}
\usepackage{lscape}

\begin{document}

\fontsize{12}{6mm}\selectfont
\setlength{\baselineskip}{2em}

$~$\\[.35in]
\newcommand{\dss}{\displaystyle}
\newcommand{\raro}{\rightarrow}
\newcommand{\be}{\begin{equation}}

\def\sech{\mbox{\rm sech}}
\def\sn{\mbox{\rm sn}}
\def\dn{\mbox{\rm dn}}
\thispagestyle{empty}

\begin{center}
{\Large\bf Intrinsic Formulation of Geometric}\\    [2mm]
{\Large\bf Integrability and Associated Riccati System}  \\    [2mm]
{\Large\bf Generating Conservation Laws}   \\   [2mm]
\end{center}

\vspace{1cm}
\begin{center}
{\bf Paul Bracken}                        \\
{\bf Department of Mathematics,} \\
{\bf University of Texas,} \\
{\bf Edinburg, TX  }  \\
{78541-2999}
\end{center}

\vspace{3cm}
\begin{abstract}
An intrinsic version of the integrability theorem for
the classical B\"{a}cklund theorem is presented.
It is characterized by a one-form which can be put
in the form of a Riccati system. It is shown
how this system can be linearized. Based on 
this, a procedure for generating
an infinite number of conservation laws is given.
\end{abstract}

\vspace{2mm}

\vspace{2mm}

\newpage
\begin{center}
{\bf 1. INTRODUCTION.}
\end{center}

The development of the study of nonlinear evolution
equations, their integrability and associated soliton
solutions has produced many fascinating results.
Such equations have B\"{a}cklund
transformations ${\bf [1]}$  and moreover their solutions
can be associated with the generation of such
geometrical objects as surfaces. This is the case
for both constant mean curvature surfaces as well
as surfaces which have constant Gaussian curvature ${\bf [2]}$. 
It has been shown that constant mean curvature
surfaces play an important role in soliton
theory by means of the generalized Weierstrass
representation of Konopelchenko ${\bf [2]}$. Moreover,
Sasaki ${\bf [3,4]}$ established a geometrical
interpretation for the inverse-scattering problem
which was originally formulated by Ablowitz ${\bf [5]}$ and
associates in terms of pseudospherical surfaces ${\bf [6]}$.
This encompasses a large class of nonlinear 
evolution equation. The B\"{a}cklund transformation
was originally introduced as a transformation
which maps one pseudospherical surface into
another. These ideas were developed and extended by
Chern and Tenenblat ${\bf [7]}$ who obtained a
systematic procedure to determine a linear
problem for which a given equation is the
integrability condition. They also realized how the
geometrical properties of a surface can provide
analytic input for such equations. Subsequently,
Cavalcante and Tenenblat ${\bf [8]}$ gave a method to
derive conservation laws for evolution equations
that describe pseudospherical surfaces.

it is to be understood here that in the Chern
Tenenblat approach to integrability, a type
of geometric integrability is implied.
To formulate this more precisely, 
we call an evolution equation $\Xi = 0$ geometrically
integrable if it describes a nontrivial, 
one-parameter family of pseudospherical surfaces.
There are, in addition to this, other 
formulations of integrability such as formal integrability
and the existence of a Lax pair ${\bf [9]}$ for the system. 
It is hoped that more insight
into the relationship between geometric 
integrability and other types of integrability can be 
achieved by investigating the types of 
conservation laws which can be obtained.

Thus the purpose here is to study the formulation
of B\"{a}cklund transformations intrinsically
based on a Pfaffian system for the case of nonlinear evolution equations
which describe pseudospherical surfaces. Arising out
of this, we return to the study and determination
of conservation laws for such equations. A different
procedure from that of Cavalcante and Tenenblat is applied
to yield a more standard form for these conservation laws.
How these different formulations of integrability
correspond can be seen by studying the conservation laws
which can be produced. 
Reyes ${\bf [9,10]}$ has studied these different 
types of integrability and generalized the approach
to conservation laws, but in a different form from 
that here.
The general approach is to take the Pfaffian system
which generates B\"{a}cklund transformations
and show that it can be put in Riccati form
in terms of one-forms which satisfy the structure
equations of the surface. Once this is done,
some further properties of the system will be
developed which lead to two conserved 
one-forms. It will be shown how these Riccati
equations can be transformed, or put in
correspondence with, a linear system of one-forms as well.

Finally, it will be shown how these results
can be used to generate infinite classes of
conservation laws. By introducing an appropriate
expansion for the quantity in the relevant Riccati
equation in terms of a parameter that appears
in the one-forms for the surface, an infinite
number of conservation laws can be produced.
The form of the conservation
equations which are generated by this
procedure are seen to correspond to those given by
other approaches to integrability. As an example,
the conservation laws for a specific nonlinear
equation, the MKdV equation,  will be determined at the end.

\begin{center}
{\bf 2. GEOMETRIC FORMULATION OF INTEGRABLE PFAFFIAN.}
\end{center}

Chern and Tenenblat ${\bf [6]}$ introduced
the idea of a differential equation for a
function $u (x,t)$ that describes a pseudospherical
surface. There is one such surface for each
solution, or if the equation is geometrically
integrable, there is a one-parameter family
of pseudo-spherical surfaces for each solution.
This idea of geometric integrability
can be regarded as a bridge between kinematic
and formal integrability. It will be seen
that geometrical properties of a pseudospherical
surface provide a systematic method for obtaining
an infinite number of conservation laws and often
B\"{a}cklund transformations ${\bf [11]}$. The classical
B\"{a}cklund theorem originated in the study
of pseudospherical surfaces relating solutions
of the sine-Gordon equation, and given three
solutions of the same equation, the superposition
formula provides a new solution algebraically.

Given a nonlinear differential equation for a
function $u (x,t)$, suppose operators $F$ and $G$ 
defined on an appropriate space of functions exist
such that
$$
\frac{\partial}{\partial t} F ( u (x,t))
+ \frac{\partial}{\partial x} G (u(x,t)) = 0,
\eqno(2.1)
$$
for all solutions $u$ of the initial equation.
Then (2.1) is called a conservation law of the
differential equation. The functional defined by
$$
I (u) = \int_{-\infty}^{\infty} \, F (u (x,t)) \, dx,
\eqno(2.2)
$$
is called a conserved quantity, since $I (u)$ is
independent of $t$ for each solution $u$ which
satisfies appropriate conditions as $ x \rightarrow \pm \infty$.

Let $M$ be a two-dimensional manifold, or surface,
and suppose $\{ e_1, e_2 \}$ constitute a local orthonormal frame field
and \{ $\omega_1$, $\omega_2 \}$  is the dual
coframe with $\omega_{12}$ the connection form. 
The structure equations ${\bf [12]}$ for $M$ are given by
$$
d \omega_1 = \omega_{12} \wedge \omega_2,
$$
$$
d \omega_2 = \omega_1 \wedge \omega_{12},
\eqno(2.3)
$$
$$
d \omega_{12} =- K \omega_1 \wedge \omega_2,
$$
where $K$ is the Gaussian curvature of $M$. 
Then $M$ is a pseudospherical surface when $K=-1$.

{\bf Definition 2.1} A differential equation for
$u (x,t)$ describes a pseudospherical surface
if it is a necessary and sufficient condition
for the existence of differentiable functions
$f_{\alpha \beta}$ with $1 \leq \alpha \leq 3$ and
$1 \leq \beta \leq 2$ depending on $u$ and its
derivatives such that the one forms
$$
\omega_{\alpha} = f_{\alpha 1} \, dx
+ f_{\alpha 2} \, dt,
\eqno(2.4)
$$
where $\omega_3 \equiv \omega_{12}$ satisfy
structure equations (2.3).

{\bf Theorem 2.1} Let the forms $\omega_{\alpha}$
be defined by (2.4), then the functions 
$f_{\alpha \beta}$ satisfy the following system of
equations
$$
- f_{11,t} + f_{12,x} = f_{31} f_{22} - f_{21} f_{32},
$$
$$
- f_{21,t} + f_{22,x} = f_{11} f_{32} - f_{12} f_{31},
\eqno(2.5)
$$
$$
- f_{31,t} + f_{32,x} = f_{11} f_{22} - f_{12} f_{21}.
$$
Partial derivatives may be indicated by variable subscripts.
To prove this, simply substitute one-forms (2.4) into
structure equations (2.3) and then equate the coefficients
of the corresponding two-forms on both sides.

As a consequence of Definition 2.1, each solution
of the differential equation 
provides a metric on $M$ whose Gaussian curvature 
$K =-1$. The definition assures that the equation
for $u$ is the required integrability condition.
The following Proposition will be crucial in
what follows. Due originally to Chern
and Tenenblat, a modified proof is given.

{\bf Proposition 2.1} Let $M$ be a $C^{\infty}$
Riemannian surface. $M$ is pseudospherical if and
only if given any unit vector $v_0$ tangent to
$M$ at $p_0 \in M$, there exists an orthonormal
frame field $v_1$, $v_2$ locally defined, such that
$v_1 (p_0) = v_0$ and the associated one-forms
$\theta_1$, $\theta_2$ and $\theta_{12}$ satisfy
$$
\theta_{12} + \theta_2 = 0.
\eqno(2.6)
$$
In this case, $\theta_1$ is a closed form.

Proof: It has to be shown that 
equation (2.6) is completely integrable
if and only if $M$ is a pseudospherical surface.
Let ${\cal I}$ be the ideal generated by the
form $\gamma = \theta_{12} + \theta_2$.
Then it follows from the structure equations (2.3) that
$$
d \gamma = d \theta_{12} + d \theta_2 = d \theta_{12}
+ \theta_1 \wedge \theta_2 = d \theta_{12} + \theta_1 
\wedge \gamma - \theta_1 \wedge \theta_2 = d \theta_{12}
- \theta_1 \wedge \theta_2,
$$
modulo ${\cal I}$. Therefore, ${\cal I}$ is closed
under exterior differentiation if and only  if $M$
is a pseudospherical surface. Hence, the first part
follows from the Frobenius Theorem.
The fact that $\theta_1$ is closed follows from
(2.6) upon exterior differentiation and using the 
structure equations 
$$
d \theta_1 = \theta_{12} \wedge \theta_2 = - \theta_2 
\wedge \theta_2 = 0.
$$
This is the intrinsic version of the integrability
theorem for the classical B\"{a}cklund Theorem.
It follows that the integral curves of $v_1$ and $v_2$
are geodesics and horocycles of $M$. $\clubsuit$

Now it is necessary to present an analytic form
of Proposition 2.1. To do this, relations between
the different one-forms for different
orthonormal frames on $M$ are established.
Again $M$ is a Riemannian surface, $e_1$, $e_2$
and $v_1$, $v_2$ are two orthonormal frames
with $\omega_1$, $\omega_2$, $\omega_{12}$ and
$\theta_1$, $\theta_2$ and $\theta_{12}$ the
associated one-forms.

Consider both frames with the same orientation, then
$$
e_1 = \cos \varphi \, v_1 + \sin \varphi \, v_2, \qquad
e_2 = - \sin \varphi \, v_1 + \cos \varphi \, v_2,
\eqno(2.7)
$$
and therefore,
$$
\begin{array}{ccc}
\omega_1 = \cos \varphi \, \theta_1 + \sin \varphi \, \theta_2,
&   &
\omega_2 = - \sin \varphi \, \theta_1 + \cos \varphi \, \theta_2,  \\
   &       &      \\
   &  \omega_{12} = \theta_{12} + d \varphi,    &     \\
\end{array}
\eqno(2.8)
$$
where $\varphi$ is the angle of rotation of the frames.
The representation in which $f_{21} = \eta$, where $\eta$ is
a constant parameter which appears in $\omega_2$ of (2.4),
is considered here.

{\bf Proposition 2.2} Let $\Xi = 0$ be a differential equation
which describes a pseudospherical surface with associated
one-forms given in (2.4) with $f_{21}= \eta$. Then for each
solution $u$ of $\Xi =0$, the system of equations for
$\varphi (x,t)$, namely
$$
\begin{array}{c}
\varphi_x - f_{31} - f_{11} \sin \varphi - \eta \, \cos \varphi = 0,  \\
   \\
\varphi_t - f_{32} - f_{12} \sin \varphi - f_{22} \, \cos \varphi =0,  \\
\end{array}
\eqno(2.9)
$$
is completely integrable. Moreover, for each solution $u$ of
$\Xi = 0$ and corresponding solution $\varphi$, the one-form
$$
( f_{11} \cos \varphi - \eta \sin \varphi ) \, dx
+ ( f_{12} \cos \varphi - f_{22} \, \sin \varphi ) \, dt = 0,
\eqno(2.10)
$$
is a closed one-form.

Proof: From (2.8), we have that $\theta_{12} = \omega_{12} - d \theta$,
and we can write $\theta_1$, and $\theta_2$ in terms of
$\omega_1$, $\omega_2$ as follows
$$
\theta_1 = \cos \varphi \, \omega_1 - \sin \varphi \, \omega_2,
\qquad
\theta_2 = \sin \varphi \, \omega_1 + \cos \varphi \, \omega_2.
\eqno(2.11)
$$
From Proposition 2.1, $u$ is a solution of $\Xi = 0$ if
and only if $\theta_{12} = - \theta_2$, hence
from (2.8), $\omega_{12}$ takes the form
$$
\omega_{12} - d \varphi + \sin \varphi \, \omega_1
+ \cos \varphi \, \omega_2 = 0,
\eqno(2.12)
$$
and is completely integrable for $\varphi$.
Moreover, $\theta_1$ in (2.11) is a closed form.

By writing $d \varphi = \varphi_x \, dx + \varphi_t \, dt$
in coordinates and using (2.4), equation (2.12)
becomes 
$$
f_{31} \, dx + f_{32} \, dt - \varphi_x \, dx - \varphi_t \, dt
+ \sin \varphi f_{11} \, dx + \sin \varphi f_{12} \, dt
+ \eta \cos \varphi \, dx + \cos \varphi f_{22} \, dt =0.
$$
Equating the coefficients of $dx$ and $dt$ to zero,
we obtain the pair (2.9), with integrability 
condition $\Xi =0$. The closed form $\theta_1$ can be
written as
$$
\theta_1 = \cos \varphi ( f_{11} \, dx + f_{12} \, dt) - \sin \varphi
( \eta \, dx + f_{22} \, dt)
= ( f_{11} \cos \varphi - \eta \sin \varphi ) \, dx
+ ( f_{12} \cos \varphi - f_{22} \sin \varphi ) \, dt.
$$
$\clubsuit$

It can be shown by straightforward differentiation that
the form $\theta_1$ and the form in (2.12) are closed
one-forms as long as the $\omega_{\alpha}$ satisfy (2.2)
irrespective of what form the $\omega_{\alpha}$ take 
in coordinates. Thus (2.4) is one particular way
of expressing the $\omega_{\alpha}$.

{\bf Lemma 2.1} The differential forms 
$$
\theta_1 =
\cos \varphi \, \omega_1 - \sin \varphi \, \omega_2,
\qquad 
\Phi = \omega_{12} - d \varphi + \sin \varphi \, \omega_1 
+ \cos \varphi \, \omega_2
$$ 
are closed one-forms modulo (2.3),

The proof of complete integrability of the $\varphi$ system
when $\omega_{\alpha}$ are given by (2.4) can be formulated as follows.

{\bf Theorem 2.2} Let $f_{\alpha \beta}$, with $1  \leq \alpha \leq 3$
and $1 \leq \beta \leq 2$ be differentiable functions
of $(x,t)$ which satisfy system (2.5). Then the system
$$
\varphi_x = f_{31} + f_{11} \sin \varphi + f_{21} \cos \varphi,
\qquad
\varphi_t = f_{32} + f_{12} \sin \varphi + f_{22}\cos \varphi,
\eqno(2.13)
$$
is completely integrable for $\varphi$.

Proof: From (2.13), we calculate both derivatives $\varphi_{xt}$
and $\varphi_{tx}$, then subtract to obtain,
$$
\varphi_{xt} - \varphi_{tx} = f_{32,x} - f_{31,t}
+ ( f_{12,x} - f_{11,t}) \sin \varphi
+ ( f_{12} f_{31} - f_{11} f_{32}) \cos \varphi
+ f_{12} f_{21} \cos^2 \varphi - f_{11} f_{22} \cos^2 \varphi
$$
$$
+ ( f_{22,x} - f_{21,x}) \cos \varphi - 
( f_{22} f_{31} - f_{21} f_{32} ) \sin \varphi
- ( f_{22} f_{11} - f_{21} f_{12} ) \sin^2 \varphi .
$$
Substituting system (2.5) into this, it is found
that the right-hand side vanishes, so $\varphi_{xt}
= \varphi_{tx}$. $\clubsuit$

In certain cases, (2.9) provides B\"{a}cklund transformations
${\bf [13,14]}$ for the equation $\Xi = 0$. Suppose it is possible to
eliminate $u$ from (2.9), then an expression of the form
$$
u = H ( \varphi ),
\eqno(2.14)
$$
is obtained, as well as a differential equation for $\varphi$,
$$
L ( \varphi ) = 0.
\eqno(2.15)
$$
These last two equations (2.14) and (2.15) are
equivalent to (2.9). Thus, Proposition 2.3 follows from 
Proposition 2.2.

{\bf Proposition 2.3} Let $\Xi = 0$ be a differential
equation which describes pseudospherical surfaces with 
associated one-forms given by (2.4). Suppose (2.9) is
equivalent to a system of equations (2.14) and (2.15).
Given a solution $u$ of $\Xi =0$, the set of equations (2.9)
is completely integrable and $\varphi$ is a solution of
(2.15). Conversely, if $\varphi$ is a solution of (2.15), then $u$,
which is defined by (2.14), is a solution of the equation $\Xi = 0$.

\begin{center}
{\bf 3. ASSOCIATED RICCATI SYSTEM AND ITS LINEARIZATION.}
\end{center}

It is shown that the one-form (2.12) can be put 
in Riccati form. In this section, it is not required that the
forms $\omega_{\alpha}$ be given as in (2.4). However they must
satisfy the structure equations in (2.3).

{\bf Proposition 3.1.} $(i)$ Let $\Xi =0$ be a differential equation
describing pseudospherical surfaces with associated one-forms
$\{ \omega_1, \omega_2 , \omega_{12} \}$. Under a change
of variable $\Gamma = \tan ( \varphi / 2)$, the completely
integrable Pfaffian system (2.12) and the  closed one-form
$\theta_1$ can be expressed as
$$
2 d \Gamma = \omega_{12} + \omega_2 + 2 \Gamma \omega_1 +
\Gamma^2 ( \omega_{12} - \omega_{2}),
\eqno(3.1)
$$
$$
\Theta_1 = \omega_1 + \Gamma ( \omega_{12} - \omega_2 ).
\eqno(3.2)
$$

$(ii)$ Let $\Xi =0$ be a differential equation describing
pseudospherical surfaces with associated one-forms
$\{ \omega_1, \omega_2 , \omega_{12} \}$. Under the 
change of variable $\hat{\Gamma} = \cot (\varphi /2)$,
the completely integrable Pfaffian system (2.12) and
closed one-form $\theta_1$ can be expressed as
$$
-2 d \hat{\Gamma} = \omega_{12} - \omega_2 + 2 \hat{\Gamma} 
\omega_1 + \hat{\Gamma}^2 ( \omega_{12} + \omega_2 ),
\eqno(3.3)
$$
$$
\Theta_2 = \omega_1 + \hat{\Gamma} ( \omega_{12} + \omega_2 ).
\eqno(3.4)
$$

Proof: Since $2 d \Gamma = \sec^2 (\varphi /2) \, d \varphi$,
the Pfaffian system becomes
$$
2 d \Gamma = \sec^2 (\frac{\varphi}{2}) ( \omega_{12} + \omega_2 )
+ 2 \tan (\frac{\varphi}{2}) \omega_1 - 2 \tan^2 ( \frac{\varphi}{2}) \omega_2
$$
$$
= \omega_{12} + \omega_2 + 2 \tan ( \frac{\varphi}{2}) \omega_1
+ \tan^2 ( \frac{\varphi}{2}) ( \omega_{12} - \omega_2 )
= \omega_{12} + \omega_2 + 2 \Gamma \omega_1 + 
\Gamma^2 ( \omega_{12} - \omega_2 ).
$$
Now,
$$
\cos^2 \frac{\varphi}{2} = \frac{1}{1 + \Gamma^2},
\qquad
\sin^2 \frac{\varphi}{2} = \frac{\Gamma^2}{1 + \Gamma^2}.
$$
Substituting into $\theta_1$, we find that
$$
\theta_1 \equiv \theta_{\Gamma} =
\frac{1- \Gamma^2}{1 + \Gamma^2} \omega_1 - \frac{2 \Gamma}
{1 + \Gamma^2} \omega_2.
$$
Define the one-form $\Theta_1$ in the following way,
$$
\Theta_1 = \theta_{\Gamma} + d \ln ( 1 + \Gamma^2)
$$
$$
= \frac{1 - \Gamma^2}{1 + \Gamma^2} \omega_1
- \frac{2 \Gamma}{1 + \Gamma^2} \omega_2
+ \frac{\Gamma}{1+ \Gamma^2}
( \omega_{12} + \omega_2 + 2 \Gamma \omega_1
+ \Gamma^2 ( \omega_{12} - \omega_2 ))
$$
$$
= \omega_1 - \Gamma \omega_2 + \Gamma \omega_{12}
= \omega_1 + \Gamma ( \omega_{12} - \omega_2 ).
$$
This finishes $(i)$. The proof of $(ii)$ proceeds
in the same way starting with the definition of $\hat{\Gamma}$.

{\bf Proposition 3.2.} Define the following two completely
integrable Pfaffian equations in terms of the set of
one-forms $\omega_1$, $\omega_2$ and $\omega_{12}$ which
satisfy structure equations (2.3),
$$
0 = \gamma_1 = -2 d \Gamma + \omega_{12} + \omega_2
+ 2 \Gamma \omega_1 + \Gamma^2 ( \omega_{12} - \omega_2 ),
\eqno(3.5)
$$
$$
0 = \gamma_2 = 2 d \hat{\Gamma} + \omega_{12} - \omega_2
+ 2 \hat{\Gamma} \omega_1 + \hat{\Gamma}^2 (\omega_{12}
+ \omega_2 ).
\eqno(3.6)
$$
Then the one-forms $\gamma_1$ and $\gamma_2$ satisfy
the following pair of equations
$$
d \gamma_1 = - \gamma_1 \wedge ( \omega_1 + \Gamma 
( \omega_{12} - \omega_2 )),
\eqno(3.7)
$$
$$
d \gamma_2 = \gamma_2 \wedge ( \omega_1 + \hat{\Gamma}
( \omega_{12} + \omega_2 )),
\eqno(3.8)
$$
in which the one-forms $\Theta_1 = \omega_1 + \Gamma 
( \omega_{12} - \omega_2 )$ and $\Theta_2 = \omega_1
+ \hat{\Gamma} ( \omega_{12} + \omega_2)$ are closed one-forms.

Proof: Differentiating $\gamma_1$, we obtain upon using (2.3)
$$
d \gamma_1 = d \omega_{12} + d \omega_2 + 2 d \Gamma \wedge \omega_1
+ 2 \Gamma d \omega_1 + 2 \Gamma \, d \Gamma \wedge
( \omega_{12} - \omega_2 ) + \Gamma^2 ( d \omega_{12} - d \omega_2 )
$$
$$
= \omega_1 \wedge \omega_2 + \omega_1 \wedge \omega_{12} 
+ 2 d \Gamma \wedge \omega_1 + 2 \Gamma d \omega_{12}
\wedge \omega_2 + 2 \Gamma \, d \Gamma \wedge ( \omega_{12} - \omega_2)
+ \Gamma^2 ( \omega_1 \wedge \omega_2 - \omega_1 \wedge \omega_{12}).
$$
Now modulo $2 d \Gamma = - \gamma_1 + \omega_{12} + \omega_2
+2 \Gamma \omega_1 + \Gamma^2 ( \omega_{12} - \omega_2)$, we obtain
after simplifying
$$
d \gamma_1 = \omega_1 \wedge \omega_2 + \omega_1 \wedge \omega_{12}
+ (- \gamma_1 + \omega_{12} + \omega_2 + 2 \Gamma \omega_1 + \Gamma^2
( \omega_{12} - \omega_2)) \wedge \omega_1 + 2 \Gamma \omega_{12}
\wedge \omega_2
$$
$$
+ \Gamma ( - \gamma_1 + \omega_{12} + \omega_2 + 2 \Gamma \omega_1
+ \Gamma^2 ( \omega_{12} - \omega_2)) \wedge ( \omega_{12} - \omega_2 )
+ \Gamma^2 ( \omega_1 \wedge \omega_2 - \omega_1 \wedge \omega_{12})
$$
$$
= \omega_1 \wedge \omega_2 + \omega_1 \wedge \omega_{12} - \gamma_1
\wedge \omega_1 + \omega_{12} \wedge \omega_1 + \omega_2 \wedge
\omega_1 + \Gamma^2 ( \omega_{12} - \omega_2) \wedge \omega_1
+ 2 \Gamma \omega_{12} \wedge \omega_2
$$
$$
- \Gamma \gamma_1 \wedge ( \omega_{12} - \omega_2) - \Gamma
\omega_{12} \wedge \omega_2 + \Gamma \omega_2 \wedge \omega_{12}
+ 2 \Gamma^2 \omega_1 \wedge ( \omega_{12} - \omega_2)
+ \Gamma^2 \omega_1 \wedge \omega_2 - \Gamma^2 \omega_1 \wedge
\omega_{12}
$$
$$
= - \gamma_1 \wedge ( \omega_1 + \Gamma ( \omega_{12}
- \omega_2)).
$$
This is exactly statement (3.7). Equation (3.8) is obtained upon differentiating
$\gamma_2$ and using (2.3) modulo $2 d \hat{\Gamma}$. $\clubsuit$

Now it is shown that a linear problem can be formulated
which is equivalent to (3.1) and (3.3).

{\bf Proposition 3.3} Define a linear problem
$$
d \Psi = \Omega \Psi,
\eqno(3.9)
$$
by means of the matrix of one-forms $\Omega$ and $\Psi$
a matrix of zero-forms which are defined to be
$$
\Omega = \frac{1}{2} \left(
\begin{array}{cc}
- \omega_1  & \omega_2 - \omega_{12}   \\
\omega_2 + \omega_{12}  & \omega_1     \\
\end{array}  \right),
\qquad
\Psi = \left(
\begin{array}{c}
\psi_1   \\
\psi_2   \\
\end{array}   \right).
\eqno(3.10)
$$
Then the linear problem defined by (3.9) and (3.10)
is equivalent to the nonlinear Riccati system (3.1)
and (3.3) under the rational transformations
$$
\Gamma = \frac{\psi_2}{\psi_1},
\qquad
\hat{\Gamma} = \frac{\psi_1}{\psi_2}.
\eqno(3.11)
$$

Proof: The linear system (3.9) has the pair of components
$$
d \psi_1 = - \frac{1}{2} \psi_1 \omega_1 + \frac{1}{2} 
\psi_2 ( \omega_2 - \omega_{12}),
\qquad
d \psi_2 = \frac{1}{2} \psi_1 ( \omega_2 + \omega_{12})
+ \frac{1}{2} \psi_2 \omega_1.
\eqno(3.12)
$$
Differentiating $\Gamma$ given in (3.11), then using
(3.12) we obtain
$$
2 \psi_1^2 \, d \Gamma = 2 \psi_1 \, d \psi_2 - 2 \psi_2 \, d \psi_1
$$
$$
= 2 \psi_1 ( \frac{1}{2} \psi_1 ( \omega_2 + \omega_{12}) + \frac{1}{2}
\psi_2 \omega_1 ) - 2 \psi_2 (- \frac{1}{2} \psi_1 \omega_1 
+ \frac{1}{2} \psi_2 ( \omega_2 - \omega_{12}))
$$
$$
= \psi_1^2 ( \omega_{12} + \omega_2 ) + 2 \psi_1 \psi_2 \omega_1
+ \psi_2^2 (\omega_{12} - \omega_2 ).
$$
Dividing both sides of this by $\psi_1^2$ and using the
definition of $\Gamma$ given in (3.11), we obtain equation (3.1).

Similarly, differentiating both sides of $\Hat{\Gamma}$ in 
(3.11), there results the equations
$$
2 \psi_2^2 \, d \hat{\Gamma} = 2 \psi_2 \, d \psi_1
- 2 \psi_1 \, d \psi_2
$$
$$
= 2 \psi_2 (- \frac{1}{2} \psi_1 \omega_1 + \frac{1}{2} 
\psi_2 ( \omega_2 - \omega_{21})) - 2 \psi_1
( \frac{1}{2} \psi_1 ( \omega_{12} + \omega_2) + \frac{1}{2}
\psi_2 \omega_1)
$$
$$
= \psi_2^2 ( \omega_2 - \omega_{12}) -2 \psi_1 \psi_2
\omega_{1} - \psi_1^2 ( \omega_{12} + \omega_2).
$$
Dividing both sides of this by $\psi_2^2$, the second
Riccati equation (3.3) appears. Thus the
Riccati system (3.1) and (3.3) is equivalent to
the linear system (3.9) and (3.10). $\clubsuit$

\begin{center}
{\bf 4. CONSERVATION LAWS.}
\end{center}

The one-forms $\Theta_1$ and $\Theta_2$ which appeared
in Proposition 3.1 can be thought of as conservation
laws in some sense. In fact, the Riccati system derived there can be put
in explicit conservation law form, and to do so, we will
write the one-forms $\omega_{\alpha}$ in terms
of coordinates in a different way from that of (2.4).
In doing so, the Riccati system of equations will take a form
which is very convenient from the point of view
of obtaining and expressing the conservation laws.

Let the system of one-forms be written as
$$
\omega_1 =- \eta \, dx - 2 A \, dt,
\qquad
\omega_2 = (r+q) \, dx + (C+B) \, dt,
\qquad
\omega_{12} = (r-q) \, dx + (C - B) \, dt.
\eqno(4.1)
$$
In (4.1), $\eta$ is a parameter and
the other coefficients are functions of the coordinates.
Substituting (4.1) into structure equations (2.3), it
is found that the functions which appear as coefficients 
must satisfy the following system of equations
$$
A_x = q C - r B,
$$
$$
q_t = B_x + 2 A q - \eta B,
\eqno(4.2)
$$
$$
r_t = C_x - 2 A r + \eta C.
$$
This system is the analogue of (2.5) using the representation (4.1). 
In terms of the functions
which appear in (4.1), the following result assumes 
a concise form.

{\bf Proposition 4.1} Let the one-forms $\omega_{\alpha}$
be given by (4.1), then modulo (4.2), Riccati system 
(3.1) and (3.3) can be rewritten in the following
conservation law form
$$
( q \Gamma )_t = ( A + B \Gamma )_x ,
\qquad
( r \Hat{\Gamma})_t = (-A + C \hat{\Gamma} )_x. 
\eqno(4.3)
$$

Proof: Writing $d \Gamma$ in terms of coordinates
$(x,t)$ and substituting (4.1), equation (3.1)
becomes
$$
\frac{\partial \Gamma}{\partial x} = r - \eta \Gamma - q \Gamma^2,
\qquad
\frac{\partial \Gamma}{\partial t} = C - 2 A \Gamma - B \Gamma^2.
\eqno(4.4)
$$
Equations (4.4) imply that
$$
B \frac{\partial \Gamma}{\partial x} - q \frac{\partial \Gamma}{\partial t}
= Br - q C + ( 2 q A - \eta B ) \Gamma.
\eqno(4.5)
$$
Adding $- q_t \Gamma$ to both sides and using the expression
$B_x = q_t - 2 A q + \eta B$ from (4.2), equation (4.5)
takes the form
$$
- (q B)_t = Br - q C - B_x \Gamma - B \Gamma_x
= - (A + B \Gamma)_x.
\eqno(4.6)
$$
This is the first equation in (4.3)

Similarly, from (3.3), we can write
$$
\frac{\partial \hat{\Gamma}}{\partial x}
= q + \eta \hat{\Gamma} - r \hat{\Gamma}^2,
\qquad
\frac{\partial \hat{\Gamma}}{\partial t}
= B + 2 A \hat{\Gamma} - C \hat{\Gamma}^2.
\eqno(4.7)
$$
Therefore, equation (4.7) implies that
$$
C \frac{\partial \hat{\Gamma}}{\partial x}
- r \frac{\partial \hat{\Gamma}}{\partial t}
= Cq +\eta C \hat{\Gamma} - rB - 2 r A \hat{\Gamma}
= q C - r B + ( \eta C - 2 r A) \hat{\Gamma}.
\eqno(4.8)
$$
Adding $- r_t \hat{\Gamma}$ to both sides
of (4.8) and using (4.2) for $r_t$, this reduces to
$$
( r \hat{\Gamma})_t = ( -A + C \hat{\Gamma} )_x,
$$
as required. $\clubsuit$

Let us show how an infinite number of
conservation laws result from these results.
The Riccati equations for $\Gamma$ and $\hat{\Gamma}$
in the $x$-variable can be rearranged to take the
form
$$
\eta q \Gamma = r q - ( q \Gamma)^2 - q [
\frac{\partial}{\partial x} ( \frac{q \Gamma}{q} )],
\qquad
\eta r \hat{\Gamma} =-rq + ( r \hat{\Gamma})^2
+ r [ \frac{\partial}{\partial x} 
( \frac{ q \hat{\Gamma}}{q} )].
\eqno(4.9)
$$
A similar pair of equations can be obtained
for the $t$ derivatives.

Expand $q \Gamma$ into a power series
in inverse powers of $\eta$ so that
$$
q \Gamma = \sum_{n=1}^{\infty}
g_n \eta^{-n}.
\eqno(4.10)
$$
The $g_n$ are unknown at this point, however
a recursion relation can be obtained for the $g_n$ by
using (4.9). Substituting (4.10) into the
$\Gamma$ equation in (4.9), we find that
$$
\sum_{n=1}^{\infty} g_n \eta^{-n +1} = qr
- ( \sum_{n=1}^{\infty} g_n \eta^{-n})^2
-q \sum_{n=1}^{\infty} (\frac{g_n}{q})_x \eta^{-n}.
\eqno(4.11)
$$
Applying the Cauchy product formula to the square
in (4.11), it then takes the form
$$
g_1 + g_2 \eta^{-1} + \sum_{n=2}^{\infty} g_{n+1} \eta^{-n}
= qr - \sum_{n=2}^{\infty} ( \sum_{j=1}^{n-1} g_j
g_{n-j} ) \eta^{-n} - q (\frac{g_1}{q})_x \eta^{-1} -q
\sum_{n=2}^{\infty} ( \frac{g_n}{q})_x \eta^{-n}.
$$
Now equate powers of $\eta$ on both sides
of this expression to produce the set
of recursions,
$$
g_1 = qr,
$$
$$
g_2 =- q ( \frac{g_1}{q})_x = - q r_x,
\eqno(4.12)
$$
$$
g_{n+1} =- \sum_{k=1}^{n-1} g_k g_{n-k}
- q ( \frac{g_n}{q})_x,
\quad n \geq 2.
$$
Substituting (4.10) into (4.3), the following
system of conservation laws appears
$$
\sum_{n=1}^{\infty} \frac{\partial g_n}{\partial t}
\eta^{-n} = \frac{\partial}{\partial x} (A +  B \sum_{n=1}^{\infty}
\frac{g_n}{q} \eta^{-n} ).
\eqno(4.13)
$$
In general, $A$ and $B$ will depend on parameter $\eta$,
the function $q$ and higher derivatives of $q$. Substituting
$A$ and $B$ into (4.13) for a particular case,
(4.13) will simplify under relations (4.12), and then like
powers of $\eta$ can be equated on both sides of (4.13).
This procedure generates an infinite number of
conservation laws for the equation under examination.

To obtain conservation laws using (4.13) in a few particular
examples using this procedure, let us consider the sine-Gordon and 
MKdV systems.

1. For the sine-Gordon equation, 
$$
q = -r = u_x /2, 
\qquad
A= \frac{1}{2 \eta} \cos u,
\qquad
B=C = - \frac{1}{2 \eta} \sin u.
\eqno(4.14)
$$
Substituting (4.14) into (4.2), the first equation
in (4.2) reduces to an identity,
and the remaining two hold modulo the sine-Gordon equation
$$
u_{xt} = \sin u.
\eqno(4.15)
$$
Putting (4.14) into (4.13), it is found that
$$
\sum_{n=1}^{\infty} \frac{\partial g_n}{\partial t} \eta^{-n}
= \frac{\partial}{\partial x} ( \frac{1}{2 \eta} \cos u
- \frac{1}{2 \eta} \sin u \sum_{n=1}^{\infty}
\frac{g_n}{q} \, \eta^{-n} ).
$$
Since $g_1 =-q^2$ and $g_2 = q q_x$, the $n=1$ term
cancels on both sides given that $u (x,t)$ satisfies
(4.15), and we are left with
$$
\sum_{n=2}^{\infty} \frac{\partial g_n}{\partial t} \eta^{-n}
= - \frac{1}{2} \frac{\partial}{\partial x}
( \sin  u \, \sum_{n=2}^{\infty}
\frac{g_{n-1}}{q} \eta^{-n} ).
$$
Using $q= u_x /2$ and equating powers of $\eta$
on both sides of this, the following set of conservation laws 
results for $n \geq 2$,
$$
\frac{\partial g_n}{\partial t}
= - \frac{\partial}{\partial x} ( \frac{\sin u}{u_x} \,  g_{n-1}).
\eqno(4.16)
$$
In fact, taking $n=2$ in (4.16) reproduces (4.15) as well.

2. Consider the
case of the MKdV equation for which $r =-q$ and
$$
A= - \frac{1}{2} \eta^3 - \eta q^2,
\qquad
B =- q_{xx}  - \eta q_x - \eta^2 q - 2 q^3,
\qquad
C = q_{xx} - \eta q_x + \eta^2 q + 2 q^3.
\eqno(4.17)
$$
Putting (4.17) into (4.2), the
first equation in (4.2) reduces to an
identity, and the remaining two generate the MKdV equation
$$
q_t + 6 q^2 q_x + q_{xxx} = 0.
\eqno(4.18)
$$
Finally substituting (4.17) into (4.13), we find that
$$
\sum_{n=1}^{\infty} \frac{\partial g_n}{\partial t} \eta^{-n}
= \frac{\partial}{\partial x} ( - \eta q^2 - (\frac{q_{xx}}{q}
+ 2 q^2 ) \sum_{n=1}^{\infty} g_n \eta^{-n}
- \frac{q_x}{q} g_1 - \frac{q_x}{q} \sum_{n=1}^{\infty} g_{n+1} \eta^{-n}
- \eta g_1 - g_2 - \sum_{n=1}^{\infty} g_{n+2} \eta^{-n} ).
$$
However, from the recursions in (4.12), 
it follows that $g_1 + q^2 =0$ and 
$-q_x g_1 - q g_2 =0$. Using these to simplify this, the remaining 
coefficients of $\eta^{-n}$ can be equated on both
sides, and the following set of conservation laws
are obtained for $n \geq 1$,
$$
\frac{\partial g_n}{\partial t} = - \frac{\partial}{\partial x}
[ ( \frac{q_{xx}}{q} + 2 q^2) g_n + \frac{q_x}{q} g_{n+1}
+ g_{n+2}  ].
\eqno(4.19)
$$
It should be stated that a similar set of
equations can be developed from $\hat{\Gamma}$ based on this
Riccati system (3.3) and (4.3).

\begin{center}
{\bf 5. CONCLUSIONS.}
\end{center}

It has been shown how the integrability theorem
for the classical B\"{a}cklund Theorem can be formulated
intrinsically, and how this can lead to the 
determination of B\"{a}cklund transformations.
More importantly, a corresponding Riccati system is given 
in intrinsic form as well for which it is shown there 
exists a linearization. Finally, by taking the
one-forms for the structure equations in a
particular way, conservation laws which correspond
to those obtained other ways are obtained.

\newpage
\begin{center}
{\bf REFERENCES.}
\end{center}

\noindent
$[1]$ M. J. Ablowitz and P. A. Clarkson,
Solitons, nonlinear evolution equations and inverse scattering,
(Cambridge University Press, NY, 1991).  \\
$[2]$ P. Bracken and A. M. Grundland, J. Math. Phys.
{\bf 42}, 1250-1282, (2001).   \\
$[3]$ R. Sasaki, Nucl. Phys. {\bf B 154}, 343-357, (1979).  \\
$[4]$ M. Wadati, H. Sanuki and K. Konno, Prog. Theor. Physics,
{\bf 53}, 419-436, (1975).  \\
$[5]$ M. J. Ablowitz, D. J. Kaup, A. C. Newell and H. Segur,
Phys. Rev. Lett., {\bf 30}, 1262-1264, (1973).  \\
$[6]$ H. H. Chen, Phys. Rev. Lett., {\bf 33}, 925-928, (1974).  \\
$[7]$ S. S. Chern and K. Tenenblat, Studies  Appl. Math.,
{\bf 74}, 55-83, (1986).  \\
$[8]$ J. A. Cavalcante and K. Tenenblat, J. Math. Phys.
{\bf 29}, 1044-1049, (1988).  \\
$[9]$ E. G. Reyes, J. Math. Phys. {\bf 41}, 2968-2989, (2000).  \\
$[10]$ E. G. Reyes, Acta Appl. Mathematicae, {\bf 64},
75-109, (2000).  \\
$[11]$ C. Rogers and W. Schief, B\"{a}cklund and Darboux transformations:
geometry and modern applications in soliton theory, (Cambridge University Press,
NY, 2002).  \\
$[12]$ S. S. Chern, W. H. Chen and K. S. Lam, Lectures on Differential
Geometry, (World Scientific, Singapore, 1999).  \\
$[13]$ R. Beals and K. Tenenblat, Studies Appl. Math. {\bf 78}, 227-256, 
(1988).  \\
$[14]$ R. Beals, M. Rebelo and K. Tenenblat, Studies Appl. Math.
{\bf 81} 125-151, (1989).  \\
\end{document}